\documentclass[10pt,twocolumn,letterpaper]{article}

\usepackage[pagenumbers]{cvpr} 

\usepackage{multirow}
\usepackage{makecell}

\usepackage{wrapfig}
\usepackage{booktabs} 
\usepackage{booktabs}

\definecolor{cvprblue}{rgb}{0.21,0.49,0.74}
\usepackage[pagebackref,breaklinks,colorlinks,allcolors=cvprblue]{hyperref}

\usepackage{xcolor}

\def\paperID{*****} 
\def\confName{CVPR}
\def\confYear{2026}

\title{ScribbleEdit: A Benchmark for Scribble-Only Image Editing}

\author{Jie Ren$^{1}$\thanks{Correspondence to: rjie@mit.edu}, Hao Kang$^{2}$, Kai Guo$^{3}$, Yiding Yang$^{2}$, Bo Liu$^{2}$, Liming Jiang$^{2}$, Qing Yan$^{2}$, Zichuan Liu$^{2}$, \\ Yizhi Song$^{2}$, Yue Xing$^{3}$, Hui Liu$^{3}$, Xin Lu$^{2}$\\
\\
$^{1}$MIT~~~~ $^{2}$ByteDance~~~~ $^{3}$Michigan State University
}

\begin{document}
\maketitle
\begin{abstract}
Scribble-based interaction provides a lightweight and intuitive way for users to specify image editing intents in interactive editing tools.
However, current image editing models based on VLMs or LLMs struggle to understand and execute edits based solely on scribble inputs. To systematically study this problem, we construct a new benchmark, \textbf{ScribbleEdit}, that evaluates the ability of image editing models to perform image editing conditioned on scribbles. This task requires both a deep understanding of the intention of the scribble and an accurate interpretation of its spatial information. In ScribbleEdit, we design an automated data construction pipeline and introduce a dedicated evaluation protocol that explicitly measures intention alignment. Our analysis reveals that existing VLM/LLM-based editing models fail to accurately capture scribble intentions. To guide future progress on scribble-only image editing, we propose a simple yet effective soft-token baseline, which enhances the model’s understanding of scribble semantics and outperforms standard image editing models on our benchmark. Our evaluation and baseline together provide a concrete foundation for assessing and improving the scribble-driven image editing.
\end{abstract}

\section{Introduction}
\label{sec:intro}

\begin{figure}[t]
  \centering
   \includegraphics[width=\linewidth]{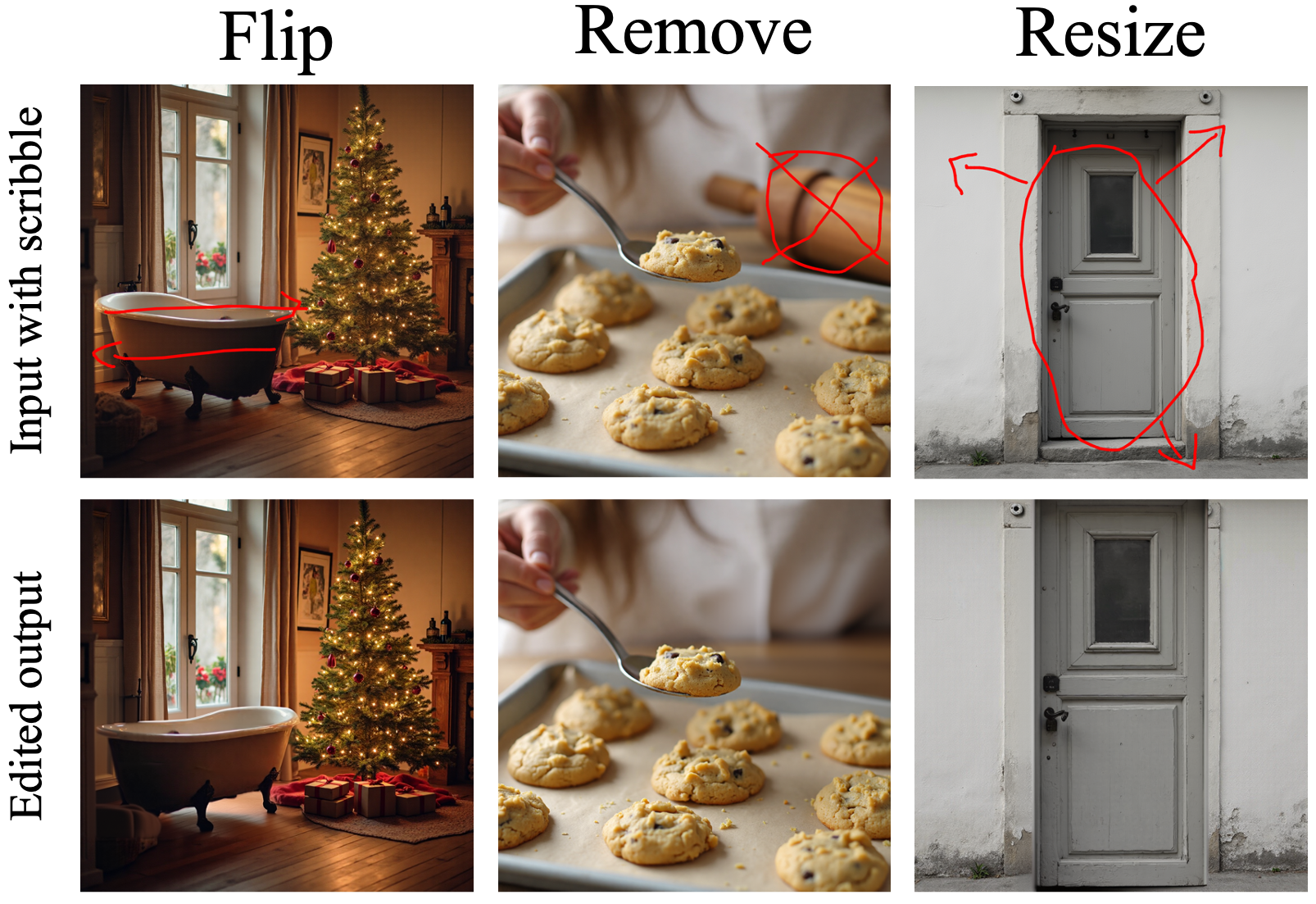}

   \caption{Illustration of image editing driven solely by user-provided scribbles. The edited results are generated by the proposed SSE (Soft-Token Scribble Editing) baseline.}
   \label{fig:intro_ex}
\end{figure}

Recently, image editing models ~\cite{couairon2022diffedit, mou2023dragondiffusion, mu2025editar, lai2025unleashing, batifol2025flux, avrahami2025stable} have demonstrated remarkable capabilities in generating high-resolution, high-quality images across a wide range of subjects. These models have been widely adopted for various downstream tasks, such as image inpainting~\cite{corneanu2024latentpaint, yu2021diverse}, image personalization~\cite{mou2025dreamo,nam2024dreammatcher,feng2025personalize} and style transfer~\cite{lei2025stylestudio, wang2025omnistyle, liu2025step1x}. Among them, natural language guided image editing has become an important conditional variant that allows users to express their intentions through textual instructions. This technique translates general user intentions into editing operations, greatly enhancing both the flexibility and diversity of image editing~\cite{deng2025emerging, bai2025qwen2, wu2025omnigen2, GeminiTeam2025}. 
However, when guidance is provided in the form of natural language, it also encounters certain limitations.
First, it cannot provide precise spatial information such as which region or object should be edited. Second, textual interaction is often tedious and cumbersome in certain application scenarios (\textit{e.g.}, when operating on mobile devices), where concise and visual inputs are preferred. As the use of image editing on mobile devices continues to grow~\cite{sargsyan2023mi, zhang2023faster}, developing more convenient and efficient guidance mechanisms for image editing has become increasingly important.

On the other hand, scribble is a suitable form of expression for this task. With just a few simple strokes, it can precisely convey the spatial information required for image editing, while also addressing the inconvenience of lengthy text input on mobile devices. \cref{fig:intro_ex} shows the examples of scribble-based image editing. However, due to the lack of dedicated datasets and benchmarks for such a task, it remains unclear whether existing general-purpose vision-language models can effectively interpret scribbles drawn on input images. Addressing this limitation is crucial, as an inability to correctly understand user scribbles would lead to inaccurate or unintended edits, ultimately undermining the usability of interactive image editing systems.

To address this gap, we introduce ScribbleEdit, a benchmark designed to evaluate image editing capabilities with scribbles as the sole input. For data, ScribbleEdit provides 23,772 synthetic scribble training image pairs and 250 manually labeled evaluation image pairs. In terms of evaluation metrics, beyond the traditional measures that assess visual similarity~\cite{hore2010image, peng2006clip} between the edited and ground-truth images, we further introduce a classification-based metric to evaluate whether a model correctly performs the operations indicated by the scribbles.
This metric explicitly separates a model’s ability to understand the intention behind the scribbles from the overall evaluation of editing quality.
With ScribbleEdit, we provide a comprehensive framework for evaluating the understanding and execution capabilities of image editing models under scribble-based guidance, aiming to bridge the gap between visual intention expression and automated editing performance.

We empirically evaluated the performance of existing image editing methods such as FLUX.1-dev~\cite{flux2024}, Qwen-Image-Edit~\cite{bai2025qwen2}, BAGEL~\cite{deng2025emerging}, Step1X~\cite{liu2025step1x}, OmniGen2~\cite{wu2025omnigen2}, Nano Banana~\cite{GeminiTeam2025}, and GPT-image-1~\cite{openai_gpt_image_1_2024} in the new benchmark, and the key observations are as follows:

\begin{itemize}
    \item \textbf{Existing models lack the ability to accurately understand or distinguish different scribbles.} Existing editing models usually employ VLMs or LLMs to process textual and visual prompts. However, based on our new evaluation metric, we observed that these models often fail to interpret the editing intention and spatial semantics conveyed by scribbles, leading to inaccurate or inconsistent editing results.
    \item \textbf{Improving the understanding of the editing intentions behind scribbles is likely the most crucial factor for this task.} Through our experiments by finetuning recent model {OmniGen2}, we found that once a model successfully interprets the meaning of the scribbles, it can perform image editing without difficulty, and the spatial information encoded in the scribbles can further enhance the editing process. This observation indicates that the key bottleneck of this task might lie in the model’s ability to understand scribbles, rather than in its generation or editing capability.
\end{itemize}


Observing the weaknesses of existing methods, to guide future progress on this task, we develop a simple yet effective baseline for comparison, called Soft-Token Scribble Editing (SSE). This approach improves the model’s ability to differentiate between editing intentions represented by various scribbles, providing a solid starting point for future research on this problem.
Specifically, we employ a CLIP~\cite{radford2021learning} module to extract the visual embeddings of the scribbles, and train a multi-layer perceptron (MLP) to transform these embeddings into a semantic space that distinguishes scribbles with different editing intentions. The transformed embedding is then appended to the input token sequence of OmniGen2 as a soft token. With this design, SSE substantially enhances the model’s ability to understand scribble intentions and improves its overall editing performance on the ScribbleEdit benchmark. 


Overall, we envision scribble-only image editing as an essential component of a broader effort to make visual editing more intuitive, efficient, and accessible. However, scribble-based interaction should be applied carefully, as it may introduce ambiguities or unintended interpretations when the model fails to correctly understand user intentions. Scribbles, while simple and efficient, can represent a wide range of semantic and spatial intentions, and misunderstanding them can lead to undesired edits. Therefore, achieving more precise and robust scribble understanding is essential for advancing this field. Current image editing models often lack the fine-grained ability to interpret the diverse semantics and spatial cues conveyed by scribbles. Our proposed ScribbleEdit benchmark and the SSE baseline provide an initial step toward this goal, offering both a standardized evaluation framework and a practical method to guide future research on scribble understanding and scribble-only image editing.





\section{Related works}
\label{sec:related}

\subsection{Image editing} 
Image editing has long been a central task in computer vision, aiming to modify existing images while preserving realism and consistency. Early approaches typically relied on pixel-level manipulation or GAN-based models such as Pix2Pix~\cite{isola2017image} and CycleGAN~\cite{zhu2017unpaired}, which enabled paired or style-based editing through supervised training. Subsequent models like StyleGAN~\cite{karras2019style} and DeepFill~\cite{yu2018generative} extended this capability to tasks such as inpainting, style transfer, and attribute editing, yet they remained limited to specific domains or rigidly defined operations.

With the rise of diffusion models and transformer-based architectures, image editing has entered a new paradigm. Methods such as SDEdit~\cite{meng2021sdedit}, InstructPix2Pix~\cite{brooks2023instructpix2pix}, Blended Diffusion~\cite{avrahami2022blended}, GLIGEN~\cite{li2023gligen}, and DragDiffusion~\cite{shi2024dragdiffusion} allow high-quality and semantically aligned editing conditioned on flexible modalities including text, masks, reference images or point-based controls. Among them, instruction-based editing has become particularly influential, where natural language instructions guide the model to perform diverse edits without explicit pixel masks. Large vision-language models such as BLIP-Diffusion~\cite{li2023blip}, EmuEdit~\cite{sheynin2024emu}, and OmniGen2~\cite{wu2025omnigen2} further integrate text encoders or multimodal decoders to align user intentions with visual outputs, significantly improving usability and generalization across open-domain images.

However, text-based instruction still faces challenges in spatial precision and interactive convenience, especially on mobile or touch-based devices. These limitations motivate the exploration of scribble-based editing, which offers a more direct and spatially grounded form of user control.

\subsection{Scribble-based generation}

Scribble-based generation provides an intuitive and lightweight interface for visual content creation, enabling users to express editing intentions through drawings rather than textual instructions. Early works explored sketch-guided or edge-based generation, where user-provided lines or contours define the structural layout of the output. Representative approaches such as SketchGAN~\cite{liu2019sketchgan}, EdgeConnect~\cite{nazeri2019edgeconnect}, ControlNet-Scribble~\cite{zhang2023sdcontrolnetscribble}, T2I-Adapter~\cite{mou2024t2i}, and SparseCtrl~\cite{guo2024sparsectrl} effectively condition diffusion models on sparse visual cues, achieving strong spatial alignment between the input structure and generated content. However, these models primarily treat scribbles as structural constraints, focusing on geometric fidelity rather than understanding the semantic intention behind each stroke.

Recent multimodal models attempt to unify textual and visual conditioning \cite{cai2024vip,wang2025object,zhao2025v2p,lin2024draw}, but the scribble modality remains underexplored. In particular, these models lack the ability to distinguish different scribble intentions such as whether a line indicates movement, resizing, or removal. In contrast, scribble-only image editing requires the model to infer user intent purely from the scribble’s visual pattern and spatial context. To systematically evaluate this ability, we propose ScribbleEdit, a benchmark designed to measure models’ understanding of scribble intentions, along with SSE, a soft-token baseline that embeds scribble semantics into the generation process. Together, they provide a foundation for advancing semantic reasoning and controllability in scribble-based generation.

\section{ScribbleEdit}


In this section, we first present the problem statement of scribble-only image editing.
Next, we describe the data construction process of ScribbleEdit in \cref{sec:data_gene}.
Finally, we introduce the proposed new evaluation metric in \cref{sec:metric}.

\subsection{Problem statement}

The key difference between scribble and text lies in the fact that scribbles can provide spatial information both conveniently and accurately. When combined with the visual understanding ability of an editing model, they allow the model to more precisely identify which region or object to modify. Therefore, unlike textual descriptions that are often more diverse and abstract, scribbles are particularly well suited for geometric editing.

In our work, we study several types of editing operations that can be effectively expressed through scribbles, including \textit{moving objects, removing objects, resizing (scaling up or down), and flipping}.

Regarding editing models, we mainly investigate models with strong understanding capabilities, especially VLM- and LLM-based editing models. For the scribble-only editing task, to make the model adaptive to this new setting, we define the setup as follows:

\begin{itemize}
    \item The editing model is provided with a \textbf{fixed system textual prompt}, which guides the editing behavior.

    \item The editing model takes the \textbf{image annotated with scribbles} as input, and may also optionally take the \textbf{scribble itself as a separate input}.
\end{itemize}

\subsection{Data generation}
\label{sec:data_gene}

\begin{figure}[t]
  \centering
   \includegraphics[width=\linewidth]{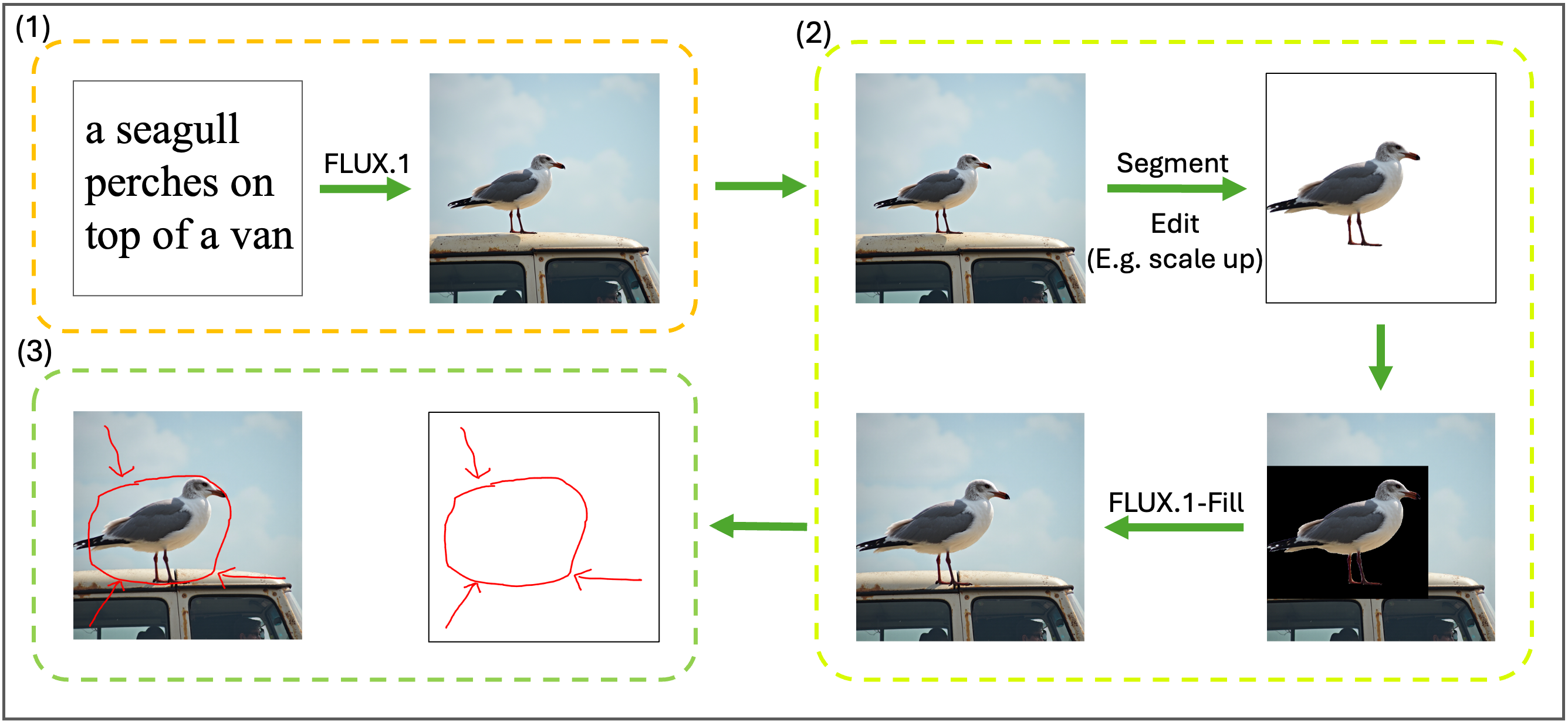}

   \caption{The pipeline of data generation}
   \label{fig:data_pipeline}
\end{figure}

To benchmark scribble-only image editing, we propose ScribbleEdit.
We provide an automated annotation pipeline capable of generating large-scale training data, as well as a set of manually annotated test samples. The pipeline and the samples together enrich the data and facilitate the development of stronger models in this line of research.
In the following, we describe the construction and design of this pipeline in detail.

As illustrated in \cref{fig:data_pipeline}, our data generation process consists of three main steps:
(1) generating candidate images,
(2) generating edited image pairs, and
(3) generating corresponding scribbles.

\textbf{Step 1. Generating candidate images.}
Existing image editing datasets typically provide image–edit pairs but are not suitable for our task. First, the editing operations in existing datasets do not cover the range of manipulations we aim to evaluate. Second, they lack the spatial annotations required for scribble-based labeling.
To address this, we first generate a collection of candidate images directly from text prompts. Specifically, we sample 10,000 prompts from the AnyEdit~\cite{yu2025anyedit} dataset and use FLUX.1-dev to synthesize 10,000 corresponding images, which serve as the initial inputs for constructing our image editing pairs.

\textbf{Step 2. Generating edited image pairs.}
To construct the corresponding edited image pairs from the candidate images, we first employ Grounded Segment Anything (GSAM)~\cite{ren2024grounded} to detect and segment objects in each image. Then, we use Llama-3-8B-Instruct~\cite{dubey2024llama} to classify the textual labels of segmented regions as foreground or background, and select the foreground objects as editing targets.
From the set of high-confidence detections, we randomly choose one object and perform spatial editing operations such as moving, resizing, or flipping. To make the edited object blend more naturally into its surroundings, we apply FLUX.1-Fill for inpainting the background around the modified area (such as the black area in \ref{fig:data_pipeline}).
Although inpainting helps smooth transitions, the edited images may still exhibit minor artifacts. To ensure higher image quality in the training set, we therefore use the original candidate images as the after-edit images (i.e. the ground truth of output), and the edited images as the before-edit counterparts.
Because different editing operations have subtle characteristics, we also design several operation-specific strategies:

\begin{itemize}
    \item \textit{Removal}. For the before-edit image, simple spatial manipulation is insufficient since removal requires the presence of an additional object that must later be deleted. To address this, we first use Qwen2.5 to propose a plausible object to be added into the scene based on the corresponding AnyEdit prompt. (For example, we can insert a ``\textit{car}'' into the image generated by the prompt ``a street sign modified to read stop bush.'') We then employ BAGEL, an image editing model, to insert this object.
    \item \textit{Moving}. For the moving operation, we randomly select a new location as the target position for the object, instead of editing in its original location.
    \item \textit{Scale-up}. Following the pipeline, one might shrink the object as the before-edit image of scale-up. However, downscaling often leaves large vacant areas that are difficult to inpaint naturally. To mitigate this, we instead use the original candidate image as the before-edit image and the enlarged (scaled-up) version as the after-edit image.
\end{itemize}

\begin{figure}[t]
  \centering
   \includegraphics[width=\linewidth]{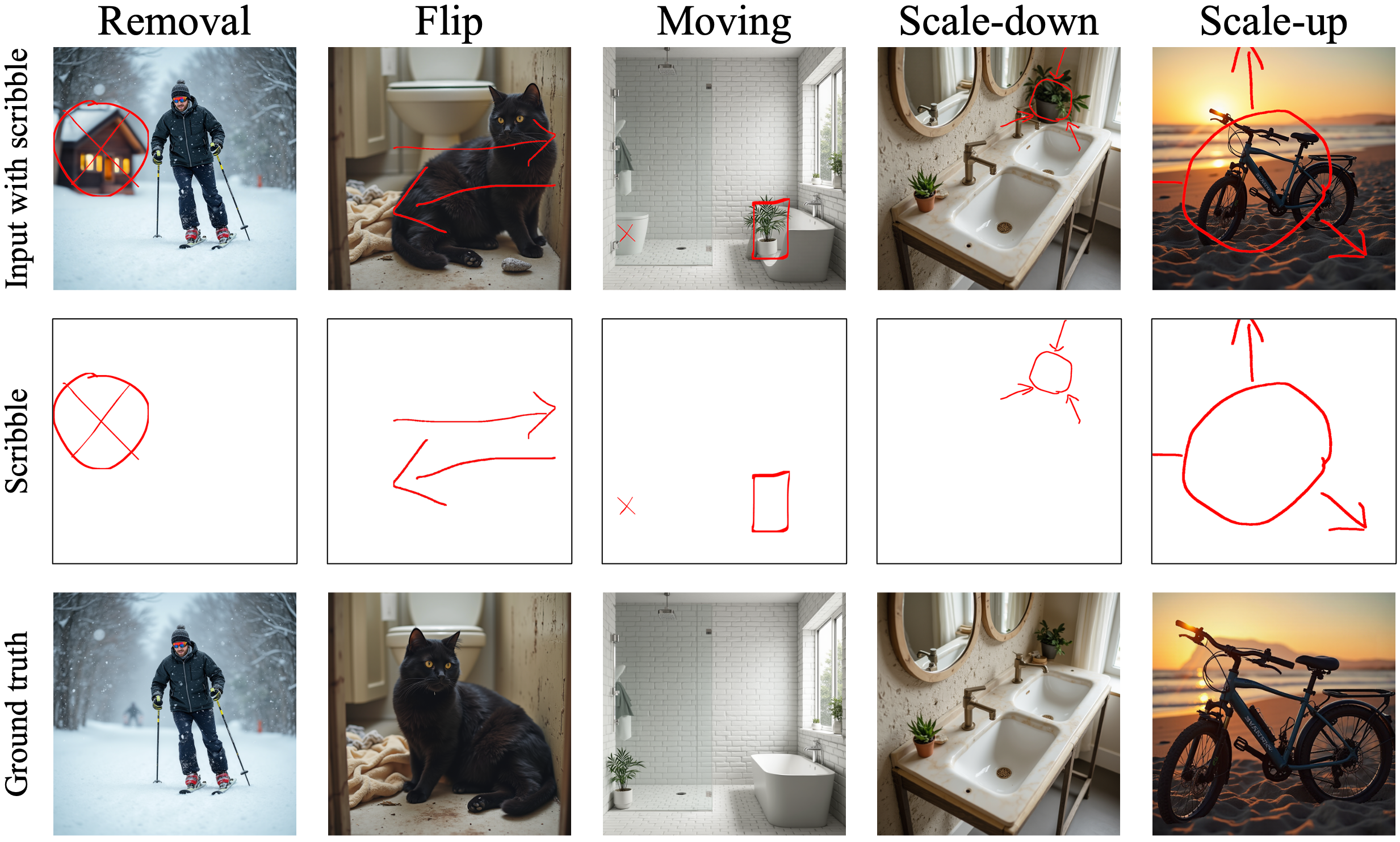}

   \caption{Example of the final generated scribble examples}
   \label{fig:scribble_ex}
\end{figure}

\textbf{Step 3. Generating scribbles.}
From the previous two steps, we have obtained both the image pairs and their corresponding edited-region masks. Based on these, we generate scribbles for different editing operations through data augmentation.
Specifically, we first manually designed a set of basic elements, including arrows, lines, circles, and rectangles. We then established composition rules to combine these primitives into scribbles that can effectively represent different operations such as remove, move, resize, and flip.
To further enhance the diversity of scribble patterns, we apply data augmentation techniques. For each primitive element, we draw multiple variants with different styles, and during composition, we randomly sample from the same category of base elements. Additionally, we apply random transformations such as rotation, resizing, and line thickening to increase diversity.
The final generated scribble examples are illustrated in \cref{fig:scribble_ex}. 

Lastly, to ensure high overall image quality, we filter out images in which the detected objects are too small to support meaningful edits. For the moving operation, we further discard samples whose source and target positions exhibit excessive spatial overlap. To balance the data distribution across operations, we double the initial pool of candidate images for the moving category.

\noindent
\textbf{Manually annotated test samples.}
From the image pairs generated in Step 2, we randomly select 50 pairs for each operation and manually annotate the corresponding scribbles on them to construct the test set.

\subsection{Metrics}
\label{sec:metric}

We introduce two metrics to comprehensively assess model performance. The first is a modified version of traditional metrics that measures overall editing quality that focuses on the edited area, and the second is a classification-based metric designed to evaluate whether the model correctly understands the editing intention conveyed by scribbles.

\begin{figure}[t]
  \centering
   \includegraphics[width=\linewidth]{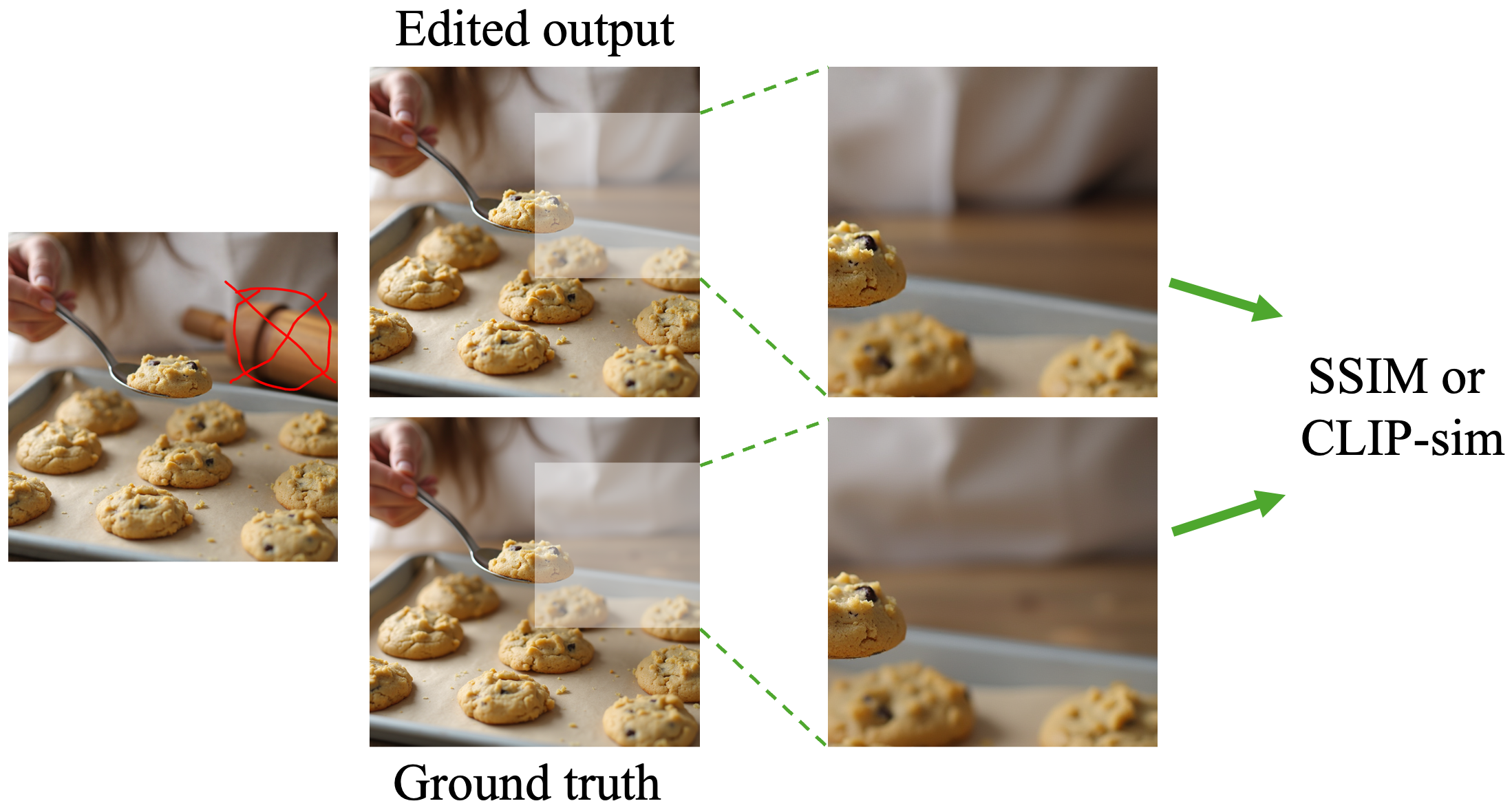}

   \caption{Focused similarity: To evaluate model performance, the edited regions of the generated outputs are compared with those of the ground truth. Specifically, the edited areas are cropped, and SSIM and CLIP-sim are computed on the cropped images.}
   \label{fig:focused_similarity}
\end{figure}

\textbf{Focused Similarity.}
A straightforward way to evaluate editing performance is to compare the model-edited image with the ground-truth image using similarity measures such as SSIM or CLIP-based semantic similarity (CLIP-sim). However, computing global image similarity can be misleading because unedited regions may dominate the metric. To address this, as shown in \cref{fig:focused_similarity}, we compute focused similarity: during data generation, we store the bounding boxes of the edited regions, and during evaluation, we crop both the generated image and the ground-truth image to these regions before computing similarity. This allows the metric to focus only on the edited area.

\noindent
\textbf{Intention Accuracy.}
To assess whether the model correctly interprets the intention behind the scribble, we train an image-pair classifier that takes the before-edit and after-edit images as input and predicts which operation has been performed.
A straightforward approach would be to train five binary classifiers. Each determines whether the editing model successfully performed flip, move, remove, scale-up, or scale-down operations. For efficiency, we instead train a single multi-class classifier that predicts which operation (or an incorrect/no operation) was applied based on the input image pair.
The “incorrect operation” category includes pairs of identical images (no edit) and their lightly augmented variants, allowing the classifier to robustly recognize both correct and failed edits. This metric provides a direct and interpretable way to evaluate whether the model has understood and executed the intended editing operation represented by the scribble.
\begin{wraptable}{r}{0.45\linewidth}  
\centering
\caption{Classification}
\label{tab:classification}
\begin{tabular}{lc}
\toprule
 & Acc. \\
\midrule
Train & 0.996 \\
Val. & 0.992 \\
Manual & 0.969 \\
\bottomrule
\end{tabular}
\end{wraptable}
We split the constructed data of ScribbleEdit into training set and validation set. As shown in \cref{tab:classification}, the accuracy on both training set are validation set are almost perfect. To test the generalization of the classifier, we further use the method introduced in the following \cref{sec:sse} to generate edited images, manually label their editing operations, and use them as test samples for the classifier.
As shown in \cref{tab:classification}, the classifier generalizes well, confirming the effectiveness of our metric.

\noindent
\section{Soft-token Scribble Editing (SSE)}
\label{sec:sse}

\begin{figure}[t]
  \centering
   \includegraphics[width=\linewidth]{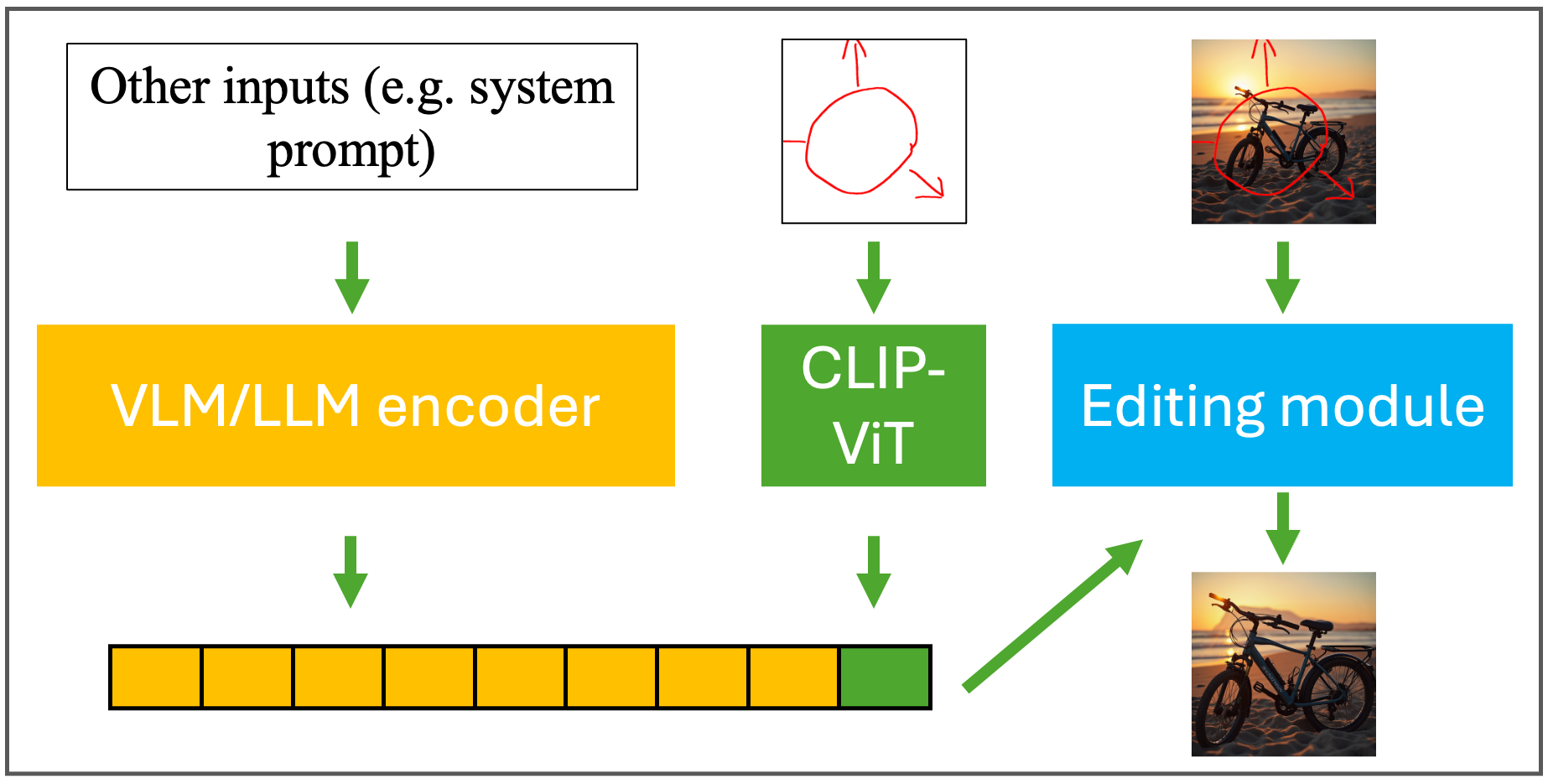}

   \caption{The framework of Soft-token Scribble Editing}
   \label{fig:sse}
\end{figure}

Based on the experiments presented in \cref{sec:exp}, we observe that current image editing models (even those built upon VLMs or LLMs with strong semantic understanding) struggle to generalize to scribble-based inputs. To guide future progress on this task, we develop a simple yet effective baseline for comparison, called Soft-Token Scribble Editing. This approach improves the model’s ability to differentiate between editing intentions represented by various scribbles, providing a solid starting point for future research on this problem.

SSE is a general framework designed for models that integrate VLMs and LLMs to process conditional inputs. In our experiments, we use OmniGen2 as a representative example.
As illustrated in \cref{fig:sse}, we employ a CLIP encoder to extract visual embeddings of the scribbles and train an MLP to project these embeddings into a semantic space that distinguishes scribbles corresponding to different editing operations.
During MLP training, we assign five randomly sampled unit vectors as target embeddings, each representing one type of operation. The MLP is optimized to produce embeddings that have high cosine similarity with their respective target vectors. Because these unit vectors are approximately orthogonal, the resulting embeddings for different operations are naturally well-separated in the semantic space.
The output embedding from the MLP is then treated as a soft token, which is appended to the end of the tokenized input sequence. Finally, we fine-tune the generation module to learn how to condition on these soft tokens to generate edits that align with the intended operation behind each scribble. In this process, we use a fixed system prompt as the input to the textual encoder, leaving the scribbles as the only variable prompt for the user interaction.

\section{Experiments}
\label{sec:exp}

\begin{table*}
  \caption{The results of Intention Accuracy, Focused SSIM and CLIP-sim on five scribble-only editing operations. The best results are highlighted in \textbf{bold}.}
  \label{tab:main}
  \centering
  \begin{tabular}{@{}l@{}lccccccccc@{}}
    \toprule
    & Operation & FLUX & Qwen & BAGEL & Step1X & OmniGen2 & Nano Banana & GPT & Fine-tuned & SSE \\
    \midrule
    \multirow{6.3}{*}{\makecell{Intention \\ Accuracy}} & Remove & 0.440 & 0.360 & 0.700 & 0.600 & 0.020 & \textbf{1.000} & 0.000 & 0.860 & {0.980} \\
    & Move & 0.100 & 0.100 & 0.040 & 0.020 & 0.040 & 0.040 & 0.021 & 0.700 & \textbf{0.920} \\
    & Flip & 0.120 & 0.020 & 0.080 & 0.000 & 0.000 & 0.200 & 0.020 & 0.840 & \textbf{0.960} \\
    & Scale up & 0.300 & 0.620 & 0.200 & 0.380 & 0.100 & 0.260 & 0.813 & 0.440 & \textbf{0.960} \\
    & Scale down & 0.180 & 0.020 & 0.300 & 0.100 & 0.960 & 0.160 & 0.560 & 0.460 & \textbf{0.860} \\
    \cmidrule{2-11}
    & Average & 0.228 & 0.224 & 0.264 & 0.220 & 0.224 & 0.329 & 0.281 & 0.660 & \textbf{0.936} \\
    \midrule
    \multirow{6.3}{*}{\makecell{Focused \\ SSIM}} & Remove & 0.326 & 0.334 & 0.370 & 0.363 & 0.250 & 0.356 & \textbf{0.398} & 0.361 & 0.395 \\
    & Move & 0.495 & 0.463 & 0.520 & 0.474 & 0.294 & 0.501 & 0.430 & 0.576 & \textbf{0.584} \\
    & Flip & 0.365 & 0.371 & 0.413 & 0.387 & 0.268 & 0.409 & 0.399 & 0.487 & \textbf{0.507} \\
    & Scale up & 0.663 & 0.683 & 0.705 & 0.731 & 0.389 & 0.762 & 0.446 & \textbf{0.891} & 0.888 \\
    & Scale down & 0.337 & 0.287 & 0.359 & 0.340 & 0.248 & 0.334 & 0.348 & 0.341 & \textbf{0.371} \\
    \cmidrule{2-11}
    & Average & 0.438 & 0.428 & 0.473 & 0.459 & 0.290 & 0.471 & 0.404 & 0.531 & \textbf{0.549} \\
    \midrule
    \multirow{6.3}{*}{\makecell{Focused \\ CLIP-sim~~}} & Remove & 0.786 & 0.831 & 0.834 & 0.856 & 0.818 & 0.872 & 0.862 & 0.893 & \textbf{0.897} \\
    & Move & 0.804 & 0.771 & 0.812 & 0.801 & 0.749 & 0.856 & 0.842 & 0.926 & \textbf{0.935} \\
    & Flip & 0.808 & 0.756 & 0.858 & 0.827 & 0.786 & 0.893 & 0.873 & 0.930 & \textbf{0.943} \\
    & Scale up & 0.869 & 0.853 & 0.898 & 0.905 & 0.840 & 0.958 & 0.894 & \textbf{0.973} & 0.969 \\
    & Scale down & 0.721 & 0.764 & 0.807 & 0.789 & 0.764 & 0.870 & 0.856 & 0.901 & \textbf{0.918} \\
    \cmidrule{2-11}
    & Average & 0.797 & 0.795 & 0.842 & 0.835 & 0.792 & 0.889 & 0.866 & 0.925 & \textbf{0.932} \\
    \bottomrule
  \end{tabular}
\end{table*}

In this section, we evaluate the performance of seven existing models and two proposed baselines.
We also analyze incorrect intention predictions, the upper bound of editing capabilities, the impact of different system textual prompts, and representative failure cases.

\subsection{Settings}

\textbf{Editing models.}
We evaluate the scribble-only editing capabilities of seven representative models, including FLUX.1-dev~\cite{flux2024}, Qwen-Image-Edit~\cite{bai2025qwen2}, BAGEL~\cite{deng2025emerging}, Step1X~\cite{liu2025step1x}, OmniGen2~\cite{wu2025omnigen2}, Nano Banana~\cite{GeminiTeam2025}, and GPT-image-1~\cite{openai_gpt_image_1_2024}. Beyond the existing models, we additionally train two variants based on OmniGen2: (1) a model fine-tuned directly on the ScribbleEdit training set, and (2) a model trained under the proposed SSE framework.

\noindent
\textbf{Metrics.}
We assess model performance using three complementary metrics. Intention accuracy measures the model’s ability to understand user intention, evaluated by a ConvNeXt classifier trained on labeled intentions. Focused similarity quantifies the consistency between the edited and reference regions using the Structural Similarity Index (SSIM) and CLIP-based semantic similarity (CLIP-sim). For CLIP-sim, we extract visual embeddings from cropped generated and reference images using a CLIP visual encoder, and compute their cosine similarity.

\noindent
\textbf{Implementation details.}
For models that accept only a single image as input, we use the scribbled image directly as input. For Nano Banana, GPT-image-1, and the SSE variant (which support multi-image conditioning) we provide both the original scribbled image and an additional image with only scribble on a white background. We include a textual system prompt describing the meaning of each scribble operation to enhance the model’s understanding of the editing intention. More details can be found in Appendix~\ref{appd:detail}.

\subsection{Main results}

In this subsection, we evaluate the performance of various models on the scribble-only image editing task using the ScribbleEdit benchmark. \cref{tab:main} presents the editing capabilities of different models across multiple editing operations. All evaluations are conducted on the test split of ScribbleEdit. For each model, we also report the average performance across all operations. The main observations are summarized as follows.

\textbf{Observation 1: Existing models perform poorly on scribble-only editing.} From \cref{tab:main}, we can observe that existing models generally perform poorly across most operations. In terms of intention understanding, even the best-performing model, Nano Banana, correctly interprets only 32.9\% of the user intentions conveyed through scribbles, while most other methods achieve merely 22–23\%. For focused SSIM, the results are mostly concentrated between 0.4 and 0.5. Although GPT-image-1 shows slightly better intention understanding, its SSIM remains lower because it tends to modify fine image details more aggressively. In contrast, CLIP-sim performs relatively better, as it focuses on semantic similarity and is less sensitive to minor texture or lighting variations. Overall, despite Nano Banana and GPT-image-1 performing slightly better than others, all existing models demonstrate unsatisfactory results.


\begin{figure*}
  \centering
  \begin{subfigure}{0.33\linewidth}
    \centering
    \includegraphics[width=.9\linewidth]{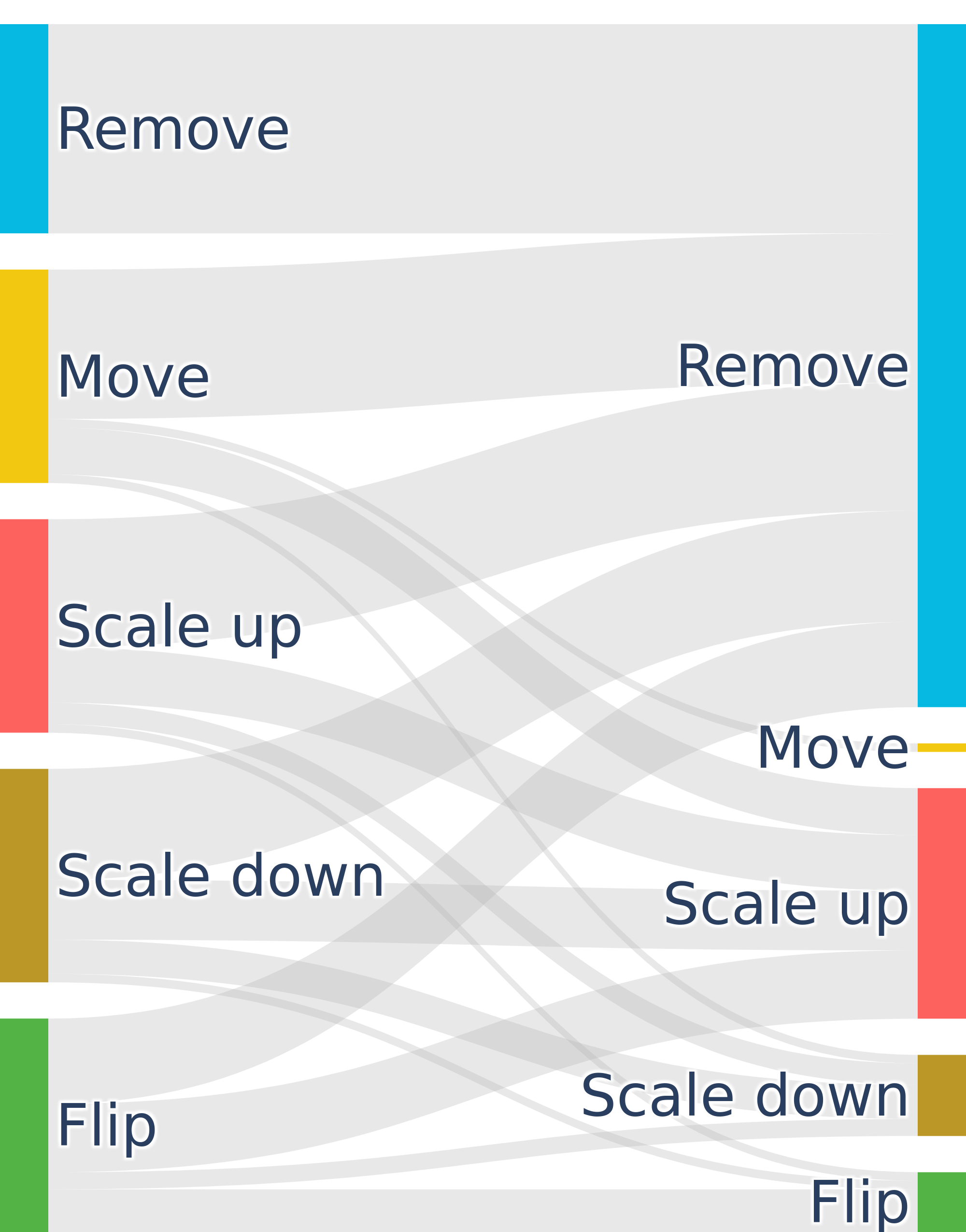}
    \caption{Nano Banana}
    \label{fig:intention_nano}
  \end{subfigure}
  \hfill
  \begin{subfigure}{0.33\linewidth}
    \centering
    \includegraphics[width=.9\linewidth]{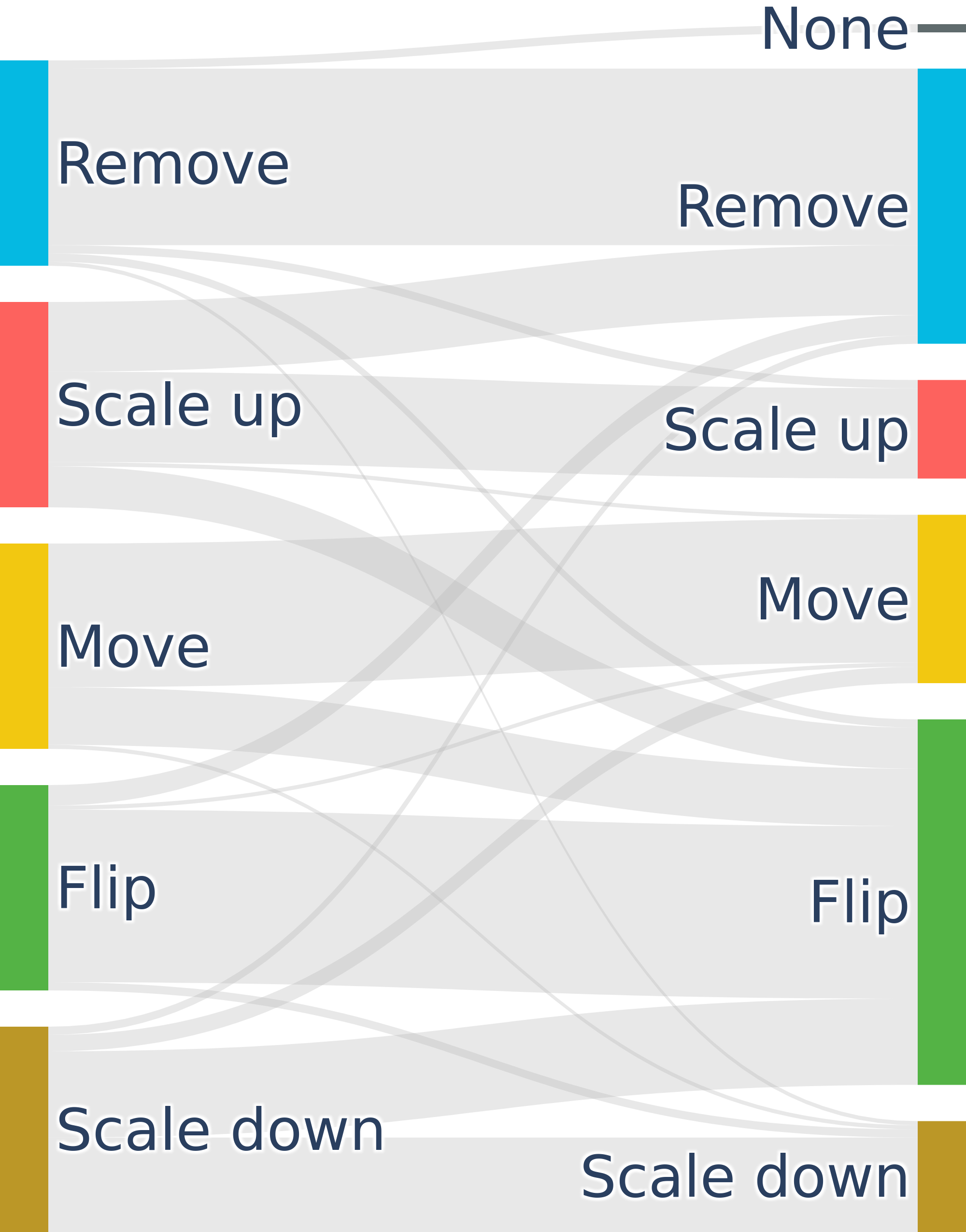}
    \caption{Fine-tuned OmniGen2}
    \label{fig:intention_base}
  \end{subfigure}
  \hfill
  \begin{subfigure}{0.33\linewidth}
    \centering
    \includegraphics[width=.9\linewidth]{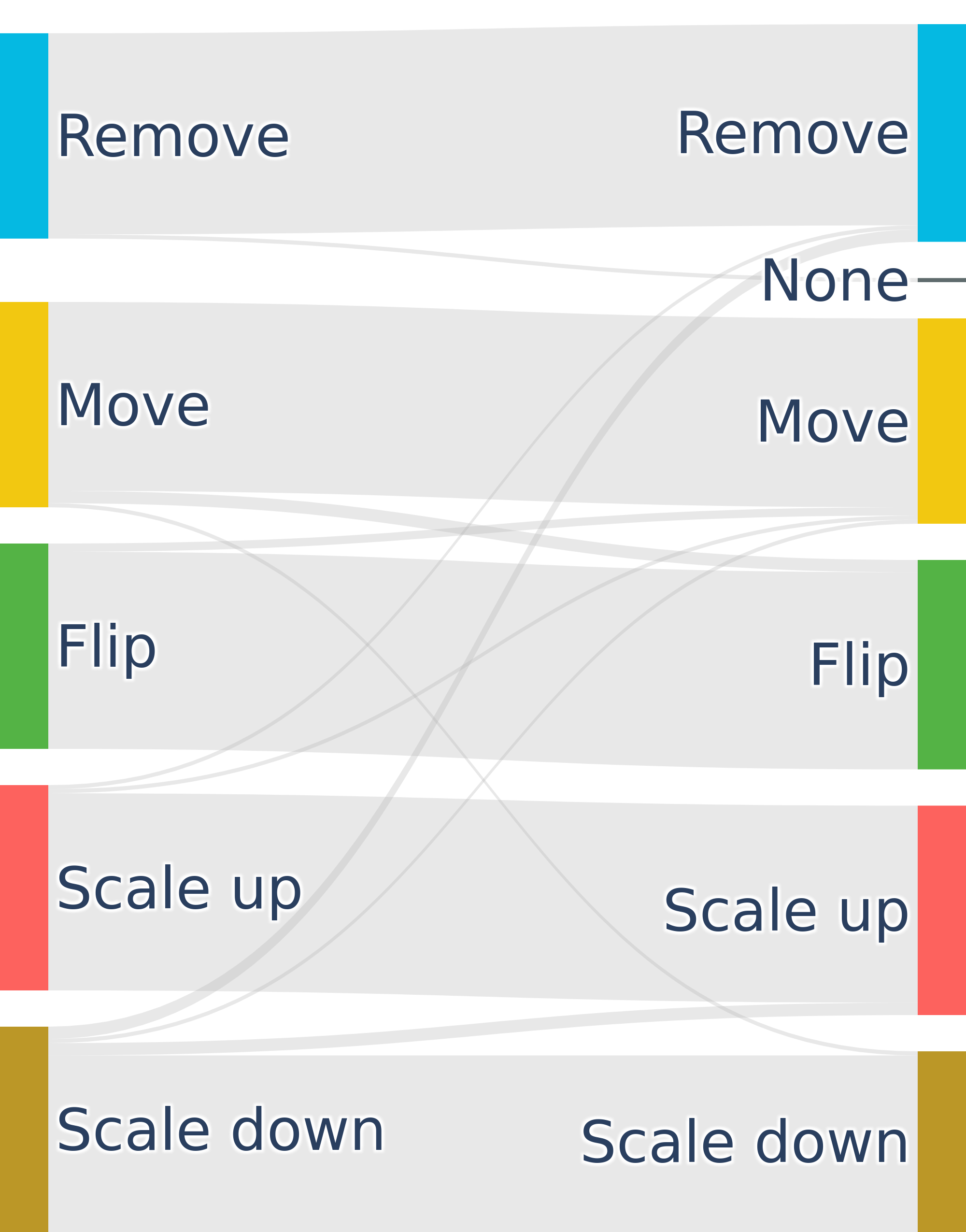}
    \caption{SSE}
    \label{fig:intention_soft}
  \end{subfigure}
  \caption{Intention understanding results of Nano Banana, the directly fine-tuned OmniGen2, and the SSE framework. For each subfigure, the left side shows the ground-truth editing intention of the scribbles, while the right side presents the intention predicted from the edited images. ``None'' means the image remains unchanged.}
  \label{fig:intention}
\end{figure*}

\textbf{Observation 2: The improvement of SSE over Fine-tuned demonstrates the importance of understanding scribbles in this task.}
Although existing models perform poorly, both the directly fine-tuned model and the SSE approach achieve promising results, with SSE performing even better. The fine-tuning procedures of these two methods are almost identical—the only difference is that SSE introduces an additional soft token which differentiates among different types of scribbles. The performance gap between the two methods suggests that understanding scribbles is a key capability missing in existing models. Once this gap is implicitly addressed through the introduction of the soft token, the results improve significantly: intention accuracy increases from an average of 66\% to 93.6\%, while focused SSIM and focused CLIP-sim rise to 0.549 and 0.932, respectively. This improvement may stem from the fact that scribbles are rarely encountered during model pre-training, even though they encode crucial information about user intentions. These findings highlight the importance of incorporating scribble-aware data and objectives in future model training.


\textbf{Observation 3: Existing models exhibit varying understanding across different operations.}
As shown in \cref{tab:main}, existing models demonstrate uneven performance across different types of editing operations. For example, removal is generally the easiest operation to understand. Except for OmniGen2 and GPT, most models achieve their highest intention accuracy on this task.
Moreover, we observe that high intention accuracy does not necessarily translate to high editing quality (i.e., high similarity with the ground truth). Although models often recognize removal intentions correctly, the resulting edited images still show low similarity to the ground truth. This is mainly because object removal requires inpainting the background area, and as discussed in \cref{sec:data_gene} (data construction), models without fine-tuning generally struggle with background inpainting. In contrast, operations such as scale up achieve better results, likely because they involve less foreground inpainting.

\subsection{Analysis of incorrect intention understanding}


In this subsection, we visualize and analyze the results of intention understanding. As shown in \cref{fig:intention}, we plot the mapping between the intended operation indicated by the input scribble and the actual operation performed by Nano Banana, the directly fine-tuned OmniGen2, and SSE. 

From the figure, we can see that Nano Banana tends to interpret a large portion of scribbles as removal, which explains why it performs well on the removal intention in \cref{tab:main}. This behavior resembles an overfitting phenomenon, where the model outputs removal regardless of the input.
In contrast, the fine-tuned OmniGen2 demonstrates a certain level of understanding of scribbles but still misclassifies many inputs as flip operations. The SSE model, however, achieves a much higher level of accuracy, correctly identifying most input intentions.

Based on our analysis of misclassified samples, we find that the causes of errors differ between Nano Banana and the fine-tuned OmniGen2. The fine-tuned OmniGen2 tends to genuinely misinterpret scribbles (especially by over-predicting flip intentions), while Nano Banana appears to lack both the ability to understand scribbles and to execute corresponding edits properly. As a result, Nano Banana often makes random modifications that the classification model subsequently interprets as removal.

Overall, these findings suggest that the classification model provides a meaningful interpretation of whether an editing model truly understands scribble intentions, especially for models trained with targeted fine-tuning, making it a useful tool for subsequent analysis.

\subsection{The upper bound of editing ability}

\begin{table}
  \caption{Upper-bound results of OmniGen2 for different editing operations. ``Single'' denotes OmniGen2 fine-tuned on data from a single operation, meaning the model does not need to distinguish between different scribble intentions. The better results between SSE and Single are highlighted in \textbf{bold}.}
  \label{tab:upper}
  \centering
  \resizebox{\linewidth}{!}{
  \begin{tabular}{lcccccc}
    \toprule
    & \multicolumn{2}{c}{\makecell{Intention \\ Accueracy}} & \multicolumn{2}{c}{\makecell{Focused \\ SSIM}} & \multicolumn{2}{c}{\makecell{Focused \\ CLIP-sim}} \\
    Operation & SSE & Single & SSE & Single & SSE & Single \\
    \midrule
    Remove & \textbf{0.980} & 0.940 & 0.395 & \textbf{0.439} & 0.897 & \textbf{0.926} \\
    Move & \textbf{0.920} & \textbf{0.920} & 0.584 & \textbf{0.604} & 0.935 & \textbf{0.944} \\
    Flip & 0.960 & \textbf{1.000} & \textbf{0.507} & 0.466 & \textbf{0.943} & 0.939 \\
    Scale up & \textbf{0.960} & \textbf{0.960} & 0.888 & \textbf{0.896} & 0.969 & \textbf{0.973} \\
    Scale down & 0.860 & \textbf{0.980} & 0.371 & \textbf{0.396} & 0.918 & \textbf{0.933} \\
    \midrule
    Average & 0.936 & \textbf{0.960} & 0.549 & \textbf{0.586} & 0.932 & \textbf{0.948} \\
    \bottomrule
  \end{tabular}
  }
\end{table}

In this subsection, we discuss the upper bound of model editing capability. Specifically, we fine-tune models using data from a single type of operation, such as training only on removal samples. In this setting, the model does not need to distinguish between different editing intentions, and the scribble only provides spatial information. We regard the performance of such a model as an approximation of the upper bound achievable through the ScribbleEdit training data. \cref{tab:upper} indicates that SSE serves as a strong baseline, achieving results close to this upper bound. This further demonstrates that enhancing a model’s understanding of scribble intentions is highly effective for this task.

\subsection{Ablations on system textual prompt}

\begin{figure}[t]
  \centering
  \resizebox{0.7\linewidth}{!}{
   \includegraphics[width=\linewidth]{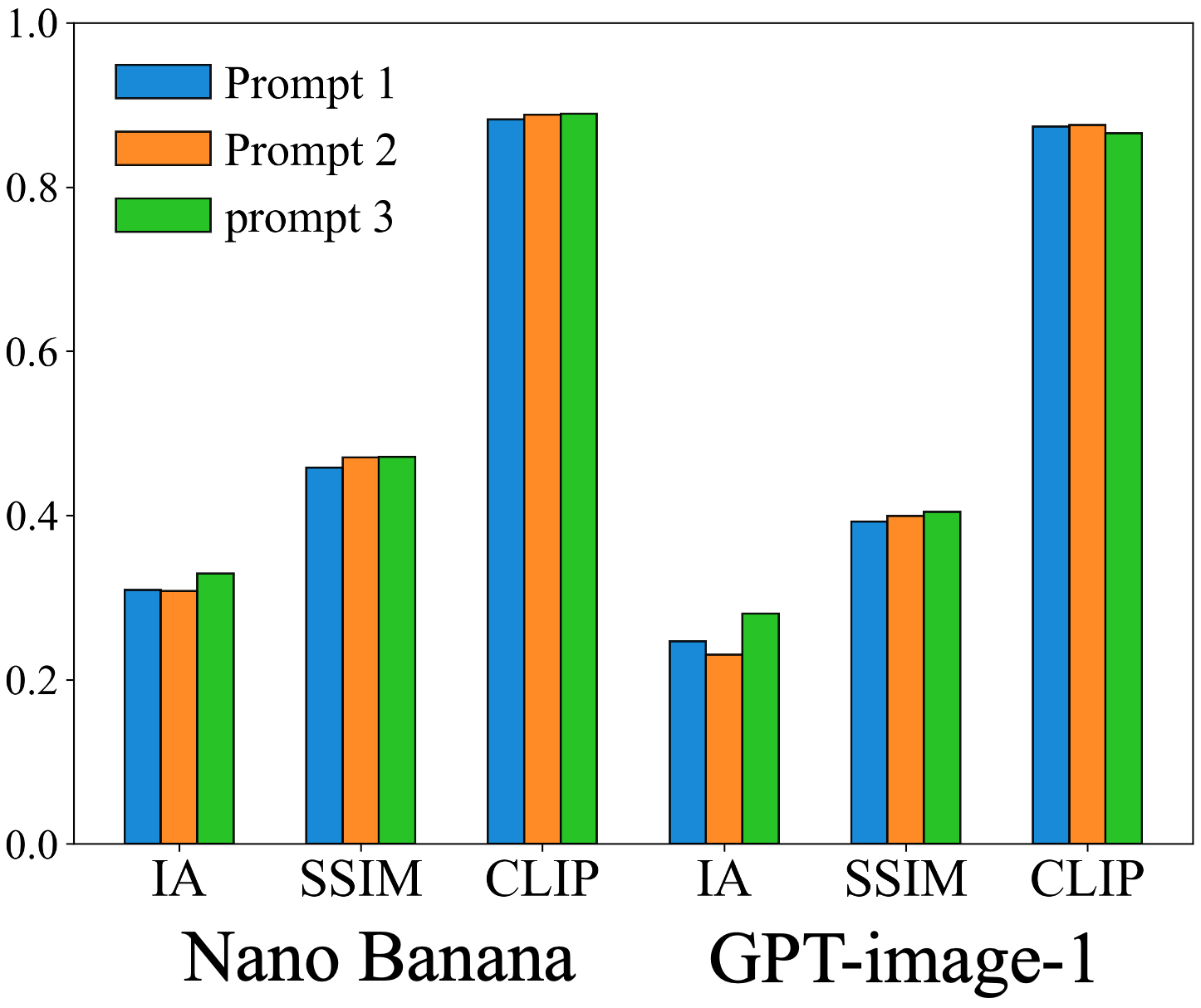}
   }

   \caption{The editing results using different prompts}
   \label{fig:textual2}
\end{figure}



In this subsection, we compare the effects of different prompts on existing models. As shown in \cref{fig:textual2}, for Nano Banana and GPT, we design three types of prompts (in the main results, we use Prompt 3, which performs best). Prompt 1 instructs the model to interpret the scribble intention and perform corresponding edits, such as removal, flip, move, and scale up or down. The image prompt includes the input image with scribbles. Prompt 2 extends this by adding textual descriptions of what each type of scribble looks like, for example, describing removal as “\textit{a circle enclosing an object with a cross over it.}” Prompt 3 further introduces an additional image showing only the scribble on a blank background. All the prompts are presented in Appendix~\ref{appd:detail}.

We find that although Prompt 3 performs slightly better than the other two in most metrics (except for GPT on CLIP-sim), the differences among the three prompts are small. This suggests that without fine-tuning, it remains difficult for models to understand scribbles, which highlights the lack of such understanding ability during pre-training and the need for further enhancement in this direction.

\subsection{Case study}

\begin{figure}[t]
  \centering
   \includegraphics[width=\linewidth]{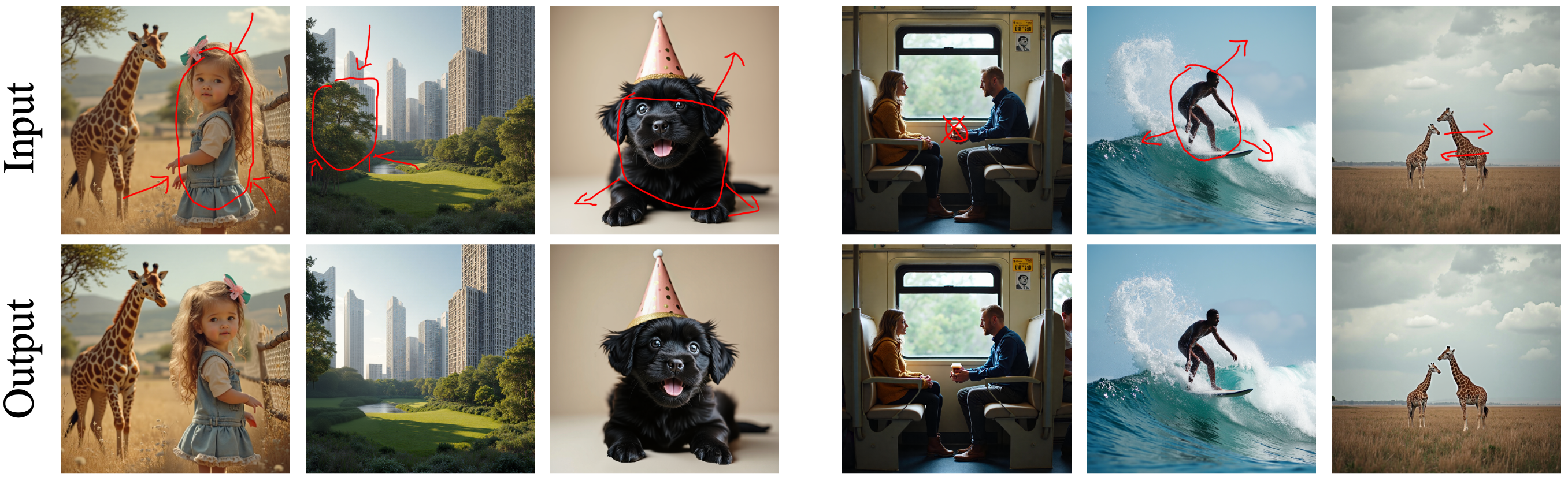}

   \caption{Two representative groups of failure cases. The first corresponds to incorrect interpretation of the scribble intention, while the second shows cases where the model failed to modify the image.}
   \label{fig:case}
\end{figure}

In this subsection, we summarize several representative cases to analyze common failure patterns in editing. We use the fine-tuned OmniGen2 as an example for discussion. As shown in \cref{fig:case}, two main types of failures can be observed. The first type involves incorrect interpretation of the scribble’s editing intention, while the second type occurs when the scribble is ignored or removed without any actual image modification. These two categories cover most of the observed failure cases. This also suggests that the difficulty of scribble-only editing lies not primarily in the editing operation itself but in the correct recognition and understanding of scribbles. This finding provides useful insights for future research in this direction.

\section{Conclusion}
\label{sec:formatting}

In this work, we introduced ScribbleEdit, a benchmark designed to systematically evaluate image editing models under scribble-only guidance. Through large-scale synthetic data generation and a dedicated evaluation metric, we revealed that existing VLM- and LLM-based models struggle to interpret the semantic and spatial intentions conveyed by scribbles. To address this challenge, we proposed the Soft-Token Scribble Editing (SSE) framework, which effectively enhances a model’s understanding of scribble intentions and significantly improves editing performance.
Our benchmark and baseline together establish a foundation for studying intention understanding and controllable editing in visual interaction. We hope this work will inspire future research toward more intuitive, efficient, and reliable image editing systems.

{
    \small
    \bibliographystyle{ieeenat_fullname}
    \bibliography{main}
}

\clearpage
\setcounter{page}{1}
\maketitlesupplementary

\section{More details}
\label{appd:detail}

\subsection{Implementation details.} 

All the models (directly fine-tuning, SSE, upper bould models) are trained for 12,000 steps with batch size of 8 following the setting of \hyperlink{https://github.com/VectorSpaceLab/OmniGen2}{https://github.com/VectorSpaceLab/OmniGen2}. 

\subsection{Prompts.} 

\begin{itemize}
    \item \textbf{Prompt 1.} \textit{``Please follow the annotations and scribbles to edit the image. The operations include object flip, resize, move, and removal.''}
    \item \textbf{Prompt 2.} \textit{``Please edit the image according to the annotations and red scribbles. The possible operations include flip, removal, move, and resize. Two horizontal arrows pointing in opposite directions (left and right) indicate a flip operation. A circle with a cross (X) drawn through it indicates removal of the object inside the circle. A rectangle enclosing an object, with a cross (X) drawn nearby, means the object should be moved to the position of the cross. A circle surrounding an object with arrows pointing toward the center means shrink the object, while arrows pointing outward from the center mean enlarge it.''}
    \item \textbf{Prompt 3.} \textit{``Please edit the image according to the annotations and red scribbles. The possible operations include flip, removal, move, and resize. Two horizontal arrows pointing in opposite directions (left and right) indicate a flip operation. A circle with a cross (X) drawn through it indicates removal of the object inside the circle. A rectangle enclosing an object, with a cross (X) drawn nearby, means the object should be moved to the position of the cross. A circle surrounding an object with arrows pointing toward the center means shrink the object, while arrows pointing outward from the center mean enlarge it. This is the scribble. \{image\} This is the image to edit with scribbles on it. \{image\}''}
    
\end{itemize}

\end{document}